# More Local Volume dwarf galaxy candidates

I. Karachentsev,[1,*] V.E. Karachentseva,[2,†] K.V. Vladimirova,[3,†]   K.A. Kozyrev[4,†]

[1]*Special Astrophysical Observatory, Russian Academy of Sciences, Nizhnii Arkhyz, 369167 Russia*
[2]*Main Astronomical Observatory, National Academy of Sciences of Ukraine, Kiev, 03143, Ukraine*
[3]*St. Petersburg University, 199034, St. Petersburg, Russia*
[4]*Kazan (Volga Region) Federal University, 420000, Kazan, Russia*

We present the results of searching for new dwarf galaxies in the Local Volume. We found 40 satellite candidates in the double-virial-radius regions of 20 Milky Way-like and LMC-like galaxies in the southern sky using DESI Legacy Imaging Surveys, 10 of which were known but not clearly associated with the Local Volume previously. Among the 40 satellite candidates, 8 are supposed members of the NGC 6744 group and 13 are located in the vicinity of the Sombrero galaxy. Based on seven companions to the giant spiral galaxy NGC 6744 with measured radial velocities, we estimate that the total mass of the group is $M_T = (1.88 \pm 0.71) \times 10^{12} M_\odot$ and the total mass-to-$K$-luminosity ratio $M_T/L_K = (16.1 \pm 6.0) M_\odot/L_\odot$. We reproduce a distribution of 68 early- and late-type galaxies in the Local Volume situated around the Sombrero, noting their strong morphological segregation and also the presence of a foreground diffuse association of dwarf galaxies at 8 degrees to SE from the Sombrero.

Ключевые слова:

## 1. INTRODUCTION

The creation of the most complete and representative sample of galaxies in the nearest volume of the universe provides an important observational basis for testing various cosmological models on small scales. This sample, limited by a distance of $\sim 10$ Mpc, was systematically created over the course of recent decades. This has been facilitated by the emergence of increasingly in-depth surveys of large sky areas in optical and radio ranges. Data on distances, radial velocities, luminosity, and other characteristics of nearby galaxies are collected in the Updated Nearby Galaxy Catalogue (UNGC), (Karachentsev et al. 2013), which is regularly updated with new objects. Currently, the Local Volume (LV) galaxy database, available at http://www.sao.ru/lv/lvgdb, contains more than 1500 galaxies. Among them, about 90% dwarf galaxies have a luminosity lower than the luminosity of the Large Magellanic Cloud (LMC).

The most recent and significant source of population growth for the LV is DESI Legacy Imaging Surveys (Dey et al. 2019), which covers about 1/3 of the entire sky. Using data from this multi-colour survey, we have found 98 new candidates to the LV members in regions of the known nearby groups (Karachentsev and Kaisina 2022, Karachentsev et al. 2023a, Karachentseva et al. 2023), as well as in the general field in direction to the Local Void (Karachentsev et al. 2023b). Most of these new dwarf galaxies were discovered in the northern hemisphere of the sky. The appearance of the tenth version of the DESI surveys (DR10) provided us the opportunity to continue searching for nearby dwarf galaxies in the southern sky. The results of our efforts are the subject of this article.

[*]Electronic address: idkarach@gmail.com
[†]Electronic address:

## 2. SEARCHING FOR DWARF SATELLITES AROUND BRIGHT SOUTHERN GALAXIES

We have not included in our program a survey of the vicinity of the very nearby galaxies: NGC 55, NGC 253, NGC 300, Cen A, and M 83, since these regions have been studied in detail by other authors using DESI survey and images obtained with large telescopes (Crnojević et al. 2019, Crosby et al. 2024a, Jones et al. 2024a, Martínez-Delgado et al. 2021, Martinez-Delgado et al. 2024, McNanna et al. 2024, Müller et al. 2015, 2024, Mutlu-Pakdil et al. 2022, Okamoto et al. 2024, Sand et al. 2024).

A list of galaxies around which we conducted our searches is presented in Table 1. Its columns contain: (1) — name of the host galaxy; (2) — equatorial coordinates of the galaxy in degrees; (3) — galaxy integral luminosity in the $K$-band, expressed in the Sun luminosity units; (4) — distance to the galaxy in Mpc; (5) — the method by which the distance was determined: TRGB — by the tip of the red giant branch, SN — by the luminosity of supernovae, TF — by the Tully-Fisher relationship (Tully et al. 2013) between the 21-cm line width and luminosity of a galaxy; (6) — the number of new candidates for satellites around the host galaxy that we have found; (7) — the size of the square in degrees within which the search for dwarf galaxies was conducted. The center of the square coincided with the host galaxy, and the square sides were oriented along RA and Dec.

For each host galaxy, we estimated the virial radius of its halo using the relation

$$(R_v/215\,\text{kpc}) = (M_T/10^{12} \times M_\odot)^{1/3}, \quad (1)$$

proposed by Tully (2015). Here, the total mass of the galaxy halo, $M_T$, was determined from its integrated $K$-band luminosity as

$$M_T/L_K = 32 \times M_\odot/L_\odot. \quad (2)$$

As Karachentsev and Kashibadze (2021) have shown, only $\sim 2/3$ of the group members are located within the virial radius, and the rest of the group population is distributed between $R_v$ and the radius of the zero-velocity sphere $R_0 \simeq 3R_v$, which separates the group volume from the general expanding field of galaxies.

Searches for new companions around large galaxies have mostly been undertaken inside the virial radius of the galaxy halo (Carlsten et al. 2022). We conducted a wider search, covering a $2R_v$ square centered on the galaxy under consideration. The median luminosity of the galaxies in our sample is $\log(L_K/L_\odot) = 9.47$, which corresponds to the LMC luminosity.

When searching for new satellites around a host galaxy, we focused on low surface brightness objects with a major angular diameter $a > 20''$. Preference was given to dwarf galaxies with a hint of granular structure and a weak brightness gradient toward the center. We also took into account the environment of the satellite candidate in order to exclude, if possible, cases of association of the dwarf with a more distant group of galaxies. For each candidate for the Local Volume objects, the maximum apparent angular diameter, apparent axial ratio, and morphological type were determined.

As a result, 40 candidate companions were detected in a total survey area of 320 square degrees around 20 host galaxies. There is a noticeable trend that the number of companions decreases rapidly with decreasing luminosity of the host galaxy. Data from deep searches for satellites conducted with large telescopes (Carlin et al. 2024, Davis et al. 2024) show that there is, on average, just under one companion per a host galaxy with a luminosity similar to the LMC luminosity.

The images of 30 candidates for new members of the Local Volume that we discovered, taken from the DESI Legacy Imaging Surveys, are shown in Figs. 1—3. The first of them contains dwarf galaxies in the vicinity of the southern host galaxies of moderate (sub-Milky Way) luminosity: NGC 628, NGC 1313, NGC 3621, NGC 5068, NGC 7090 and NGC 7424. Fig 2 and Fig 3 present, respectively, dwarf objects around the high-luminosity galaxies, NGC 6744 and NGC 4594 ("Sombrero"). In addition to them, we have noted 10 more dwarf galaxies that were known earlier with ESO, PGC, KK and KKs names, but were not considered as members of the LV. Each image size in Figs. 1—3 is $2'\times 2'$. North is up, east is left.

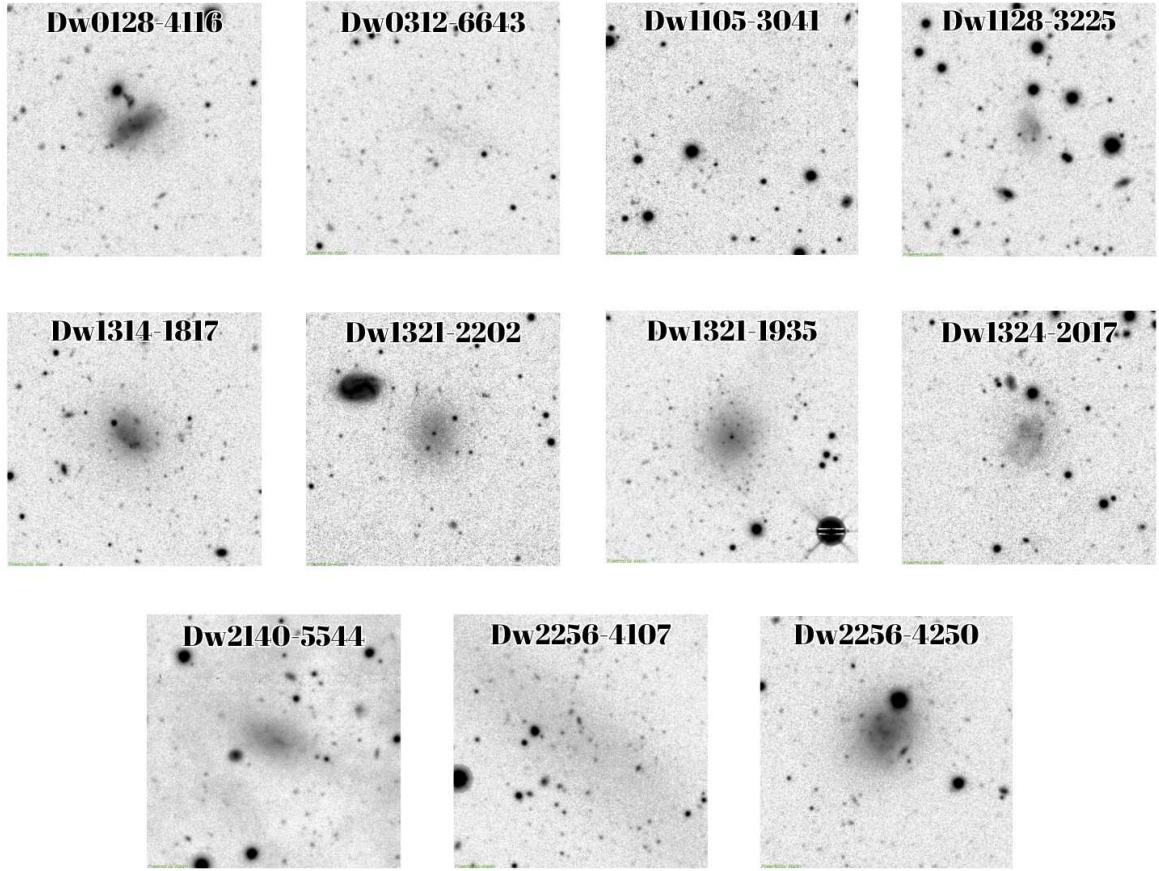

**Figure 1**: Reproduction of images of 11 nearby dwarf galaxy candidates from the DESI Legacy Imaging Surveys, found near the southern LV galaxies of sub-Milky Way luminosity. Each image size is $2'\times 2'$. North is at the top, east is on the left.

The summary list of 40 supposed dwarf satellites is given in Table 2, which contains the following data: (1) — galaxy name; (2) — equatorial coordinates; (3, 4) — angular diameter in arc minutes and apparent axial ratio of the galaxy; (5) — apparent magnitude of the galaxy in the $B$- band, determined from $g$- and $r$-magnitudes in the DESI survey as $B = g + 0.313\times(g-r) + 0.227$, or estimated visually for diffuse objects with a clump-like structure; (6) — morphological type of the dwarf galaxy: Irr — irregular, Im — Magellanic, BCD — blue compact dwarf, Sph — spheroidal, Tr — transitional between Irr and Sph. Below we note some features of the detected dwarf galaxies.

ESO 472-015. A relatively bright dwarf galaxy with a heliocentric velocity of $V_h = 655 \pm 46$ km s$^{-1}$, at a projected separation of $R_p = 90$ kpc from the host galaxy NGC 24, which has $V_h = 553\pm 2$ km s$^{-1}$ according to HyperLEDA (Makarov et al. 2014).

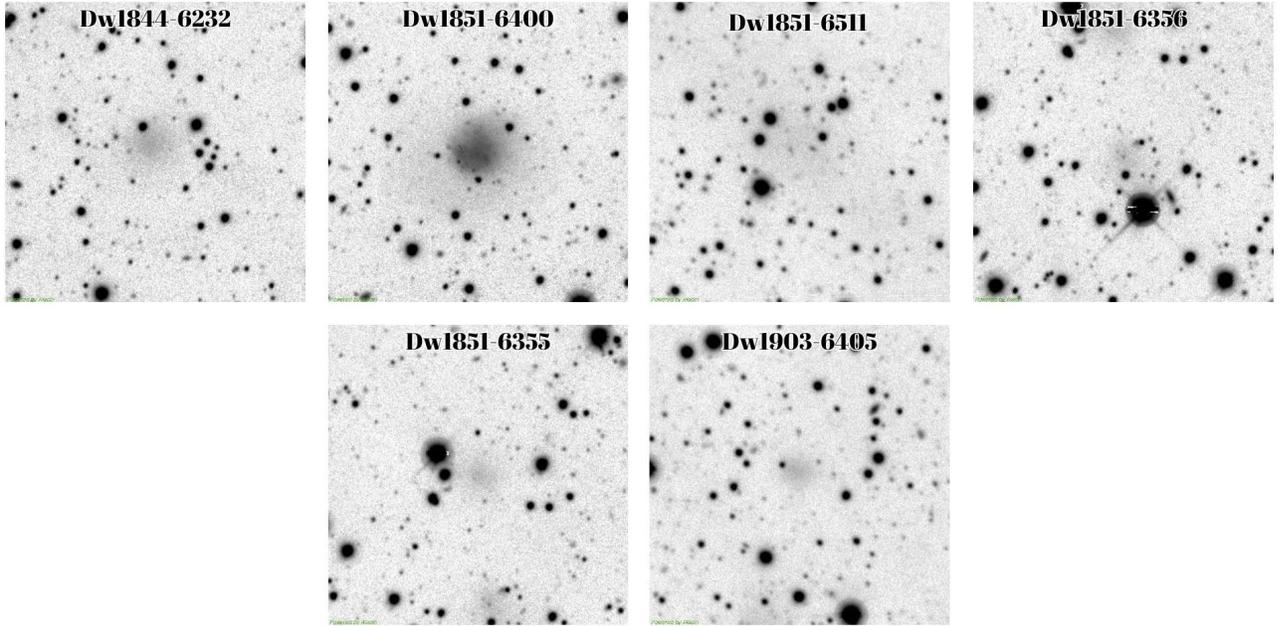

**Figure 2**: Reproduction of images of six nearby dwarf galaxy candidates from the DESI Legacy Imaging Surveys, found near the giant galaxy NGC 6744. Each image size is $2' \times 2'$. North is at the top, east is on the left.

PGC 01242. This is a BCD galaxy with a radial velocity of $V_h = 670 \pm 2$ km s$^{-1}$. The angular size is related to a faint extended halo, which is also clearly visible in the UV-band according to GALEX data (Martin et al. 2005).

Dw 0128-4116. An irregular dwarf with asymmetric halo.

Dw 0312-6643. This dwarf spheroidal galaxy of the low surface brightness and semi-resolved into stars seems to be a satellite of peculiar spiral galaxy NGC 1313.

Dw 1105-3041, KK 101, Dw 1128-3225. Probable satellites of spiral galaxy NGC 3621.

Dw 1231-1216. A dwarf galaxy with granular structure and faint asymmetric halo; a probable satellite of NGC 4594 ("Sombrero").

Dw 1245-1332. A dwarf spheroidal galaxy with a radial velocity of $V_{LG} = 783$ km s$^{-1}$ measured by Crosby et al. (2024b).

Dw 1253-0613. A dwarf spheroidal galaxy with granular structure; it belongs probably to a foreground diffuse association of dwarfs near DDO 148.

Dw 1301-1627. A dSph galaxy with granular structure, a probable member of association of dwarfs around DDO 161.

KK 186. A transition type (Tr) dwarf with a granular structure. Together with the four following objects in Table 2, it forms a retinue of the spiral galaxy NGC 5068.

Dw 1321-1935. A nucleated dSph, projected onto a very distant cluster of galaxies.

Dw 1851-6400. A Tr-type dwarf with a granular structure, a probable satellite of the giant spiral galaxy NGC 6744.

Dw 2140-5544. A spheroidal dwarf with a granular structure; located in a dense field of interstellar cirrus.

Dw 2256-4107. A satellite of the spiral galaxy NGC 7424, half-destroyed by the host.

## 3. SUITE OF DWARFS AROUND THE GALAXY NGC 6744

The Sc spiral galaxy NGC 6744 is the most extended object in the LV. Its optical Holmberg's linear diameter reaches 60 kpc, and its angular momentum is twice that of the Milky Way (Karachentsev and Zozulia 2023). Among NGC 6744's neighbours, seven galaxies in the known catalogues (NGC, IC, ESO) are associated with NGC 6744, judging by their radial velocities. Searches for fainter dwarf galaxies (Karachentseva and Karachentsev 2000) and (Karachentsev et al. 2020b) led to the discovery of three and two low surface brightness probable companions, respectively. Carlsten et al. (2022) and Hunt et al. (2024) added four more objects to the list of members of the NGC 6744 group. Carlsten et al. (2022) estimated their distances from surface brightness fluctuations (sbf). Our searches for dwarf galaxies in the wider vicinity of NGC 6744 revealed eight more new candidates for this group: Dw 1844-6232, Dw 1851-6400, Dw 1851-6511, Dw 1851-6356, Dw 1851-6355, Dw 1903-6405, KKs 73, and KKs 74. The last two of these were known previously but were not included in the LV database.

A summary list of 25 supposed members of the NGC 6744 group is presented in Table 3. Its rows contain: (1) — galaxy name; (2,3) — supergalactic coordinates in degrees; (4) — morphological type; (5) — distance to the galaxy in Mpc; (6) — method by which the distance was estimated; for two galaxies (IC 4824 and ESO 141-042), the distances were determined from their radial velocities in the Numerical Action Method (NAM) model (Kourkchi et al. 2020, Shaya et al. 2017), which takes into account the local field of peculiar velocities; (7) — projected separation of the galaxy from NGC 6744 under the assumption that all members of the group are located at a radial distance of $D = 9.51$ Mpc, measured by Anand et al. (2021) using the TRGB; (8) — radial velocity of the galaxy relative to the center of the Local Group; (9) — the total mass of the group estimated via individual members based on their projected separation $R_p$ and the difference in radial velocities $\Delta V$ relative to NGC 6744 from the relation

$$M_T/M_\odot = (16/\pi) \times G^{-1} \times \Delta V^2 \times R_p, \tag{3}$$

where $G$ is the gravitational constant. This relation assumes a chaotic orientation of the satellite orbits with an average orbital eccentricity $e = 0.7$ (Karachentsev and Kashibadze 2021).

The perimeter of our search area is shown in Fig. 4. The distribution of supposed members of the NGC 6744 group is presented in supergalactic coordinates to reduce large-scale distortions due to the convergence of meridians in the {RA, Dec} system. Previously known members of the group are shown as circles, and objects we have discovered are marked with asterisks. Spheroidal dwarf galaxies without signs of star formation are highlighted with red symbols. The radial velocities of the galaxies (in km s$^{-1}$) are indicated by numbers. The large circle centered on NGC 6744 corresponds to the virial radius of the group, 296 kpc or 1.78°.

The distribution of galaxies shows a tendency towards the presence of substructures. The asymmetric distribution of spheroidal dwarf galaxies is noteworthy: all nine Sph objects are located on the right half of the figure. We have not found an explanation for this feature. In general, such asymmetry is not uncommon among galaxy groups. There is quite an extensive literature discussing both observational manifestations of the lopsidedness of satellite galaxy systems (Brainerd and Samuels 2020, Heesters et al. 2024) and their comparison with the results of numerical simulations of the structure of groups (Pawlowski et al. 2017, Wang et al. 2021). It is also noteworthy that the distribution of potential satellites around NGC 6744 appears to be flattened in the diagonal direction, reminiscent of the issue of satellite planes seen around the Milky Way, the Andromeda nebulae and other nearby major galaxies (Ibata et al. 2013, Karachentsev and Kroupa 2024, Müller et al. 2021, Pawlowski and Kroupa 2020). The few available velocities even appear consistent with a kinematic trend: higher velocities than the host on the top-left side, lower on the bottom-right side. Future velocity measurements of the remaining supposed members of the group could easily confirm or disprove this assumption.

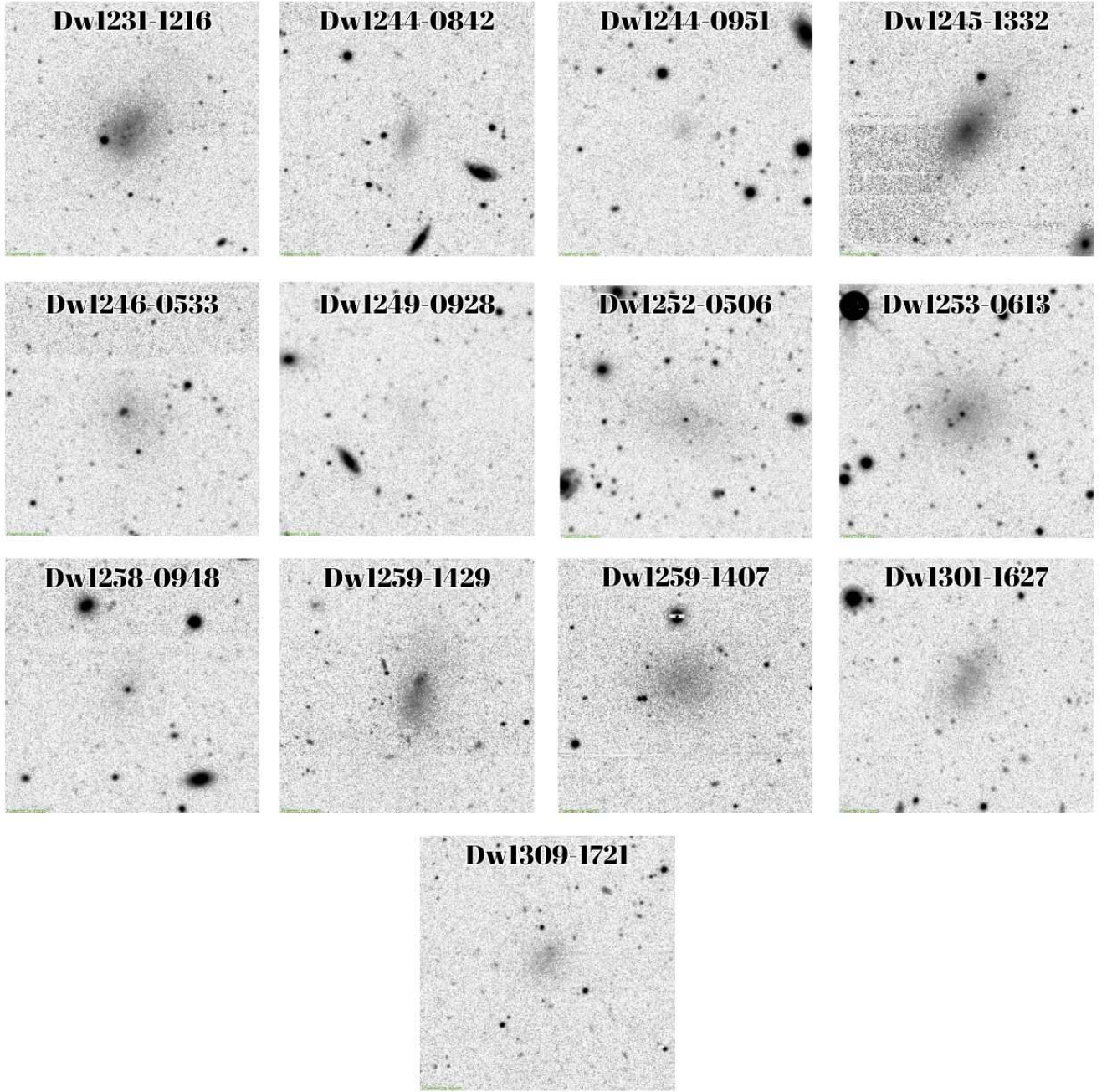

**Figure 3**: Reproduction of images of 13 nearby dwarfs from the DESI Legacy Imaging Surveys, found in the vicinity of the Sombrero galaxy. Each image size is $2' \times 2'$. North is at the top, east is on the left.

As follows from the data in the last column of Table 3, the average estimate of the total mass of the NGC 6744 group based on seven satellites is $\langle M_T \rangle = (1.88 \pm 0.71) \times 10^{12} M_\odot$. With the integrated luminosity of the group $\Sigma L_K = 1.17 \times 10^{11} L_\odot$, the ratio of the total mass to the total luminosity of the group is $(16.1 \pm 6.0) M_\odot/L_\odot$, which is two times smaller than the average ratio $(31 \pm 6) M_\odot/L_\odot$ obtained for the ensemble of 25 groups in the LV (Karachentsev and Kashibadze 2021). This difference is leveled out if we consider only the LV groups with spiral host galaxies, for which the total mass-to-K-luminosity ratio is $(17.4 \pm 2.8) M_\odot/L_\odot$ according to Karachentsev and Kashibadze (2021).

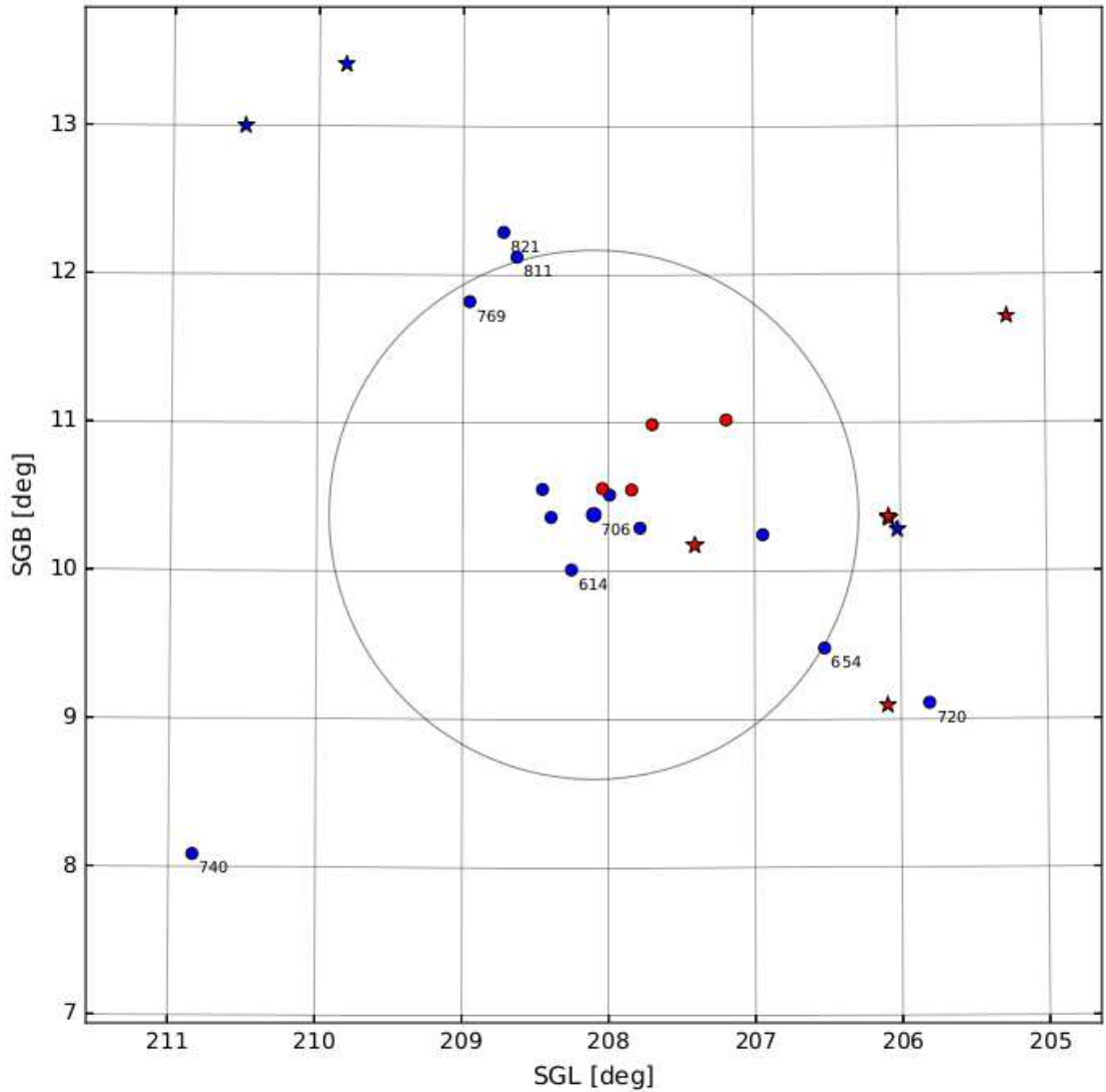

**Figure 4**: The galaxy group region around NGC 6744 in supergalactic coordinates. Previously known members of the group are shown as circles, new candidate members are shown as asterisks. Spheroidal dwarf galaxies are indicated by red symbols. The numbers indicate the radial velocity of the galaxies in km s$^{-1}$. The circle centered on NGC 6744 corresponds to the virial radius of the group of 296 kpc.

## 4. THE SURROUNDINGS OF NGC 4594

The galaxy NGC 4594 = M 104 ("Sombrero") has the highest luminosity, $\log(L_K/L_\odot) = 11.32$, among the LV population. The distance to it, $D = (9.55 \pm 0.46)$ Mpc, has been determined by McQuinn et al. (2016) via TRGB method. The search for satellites around Sombrero was undertaken by various teams (Carlsten et al. 2022, Crosby et al. 2024a, Javanmardi et al. 2016, Karachentsev et al. 2000, 2020b).

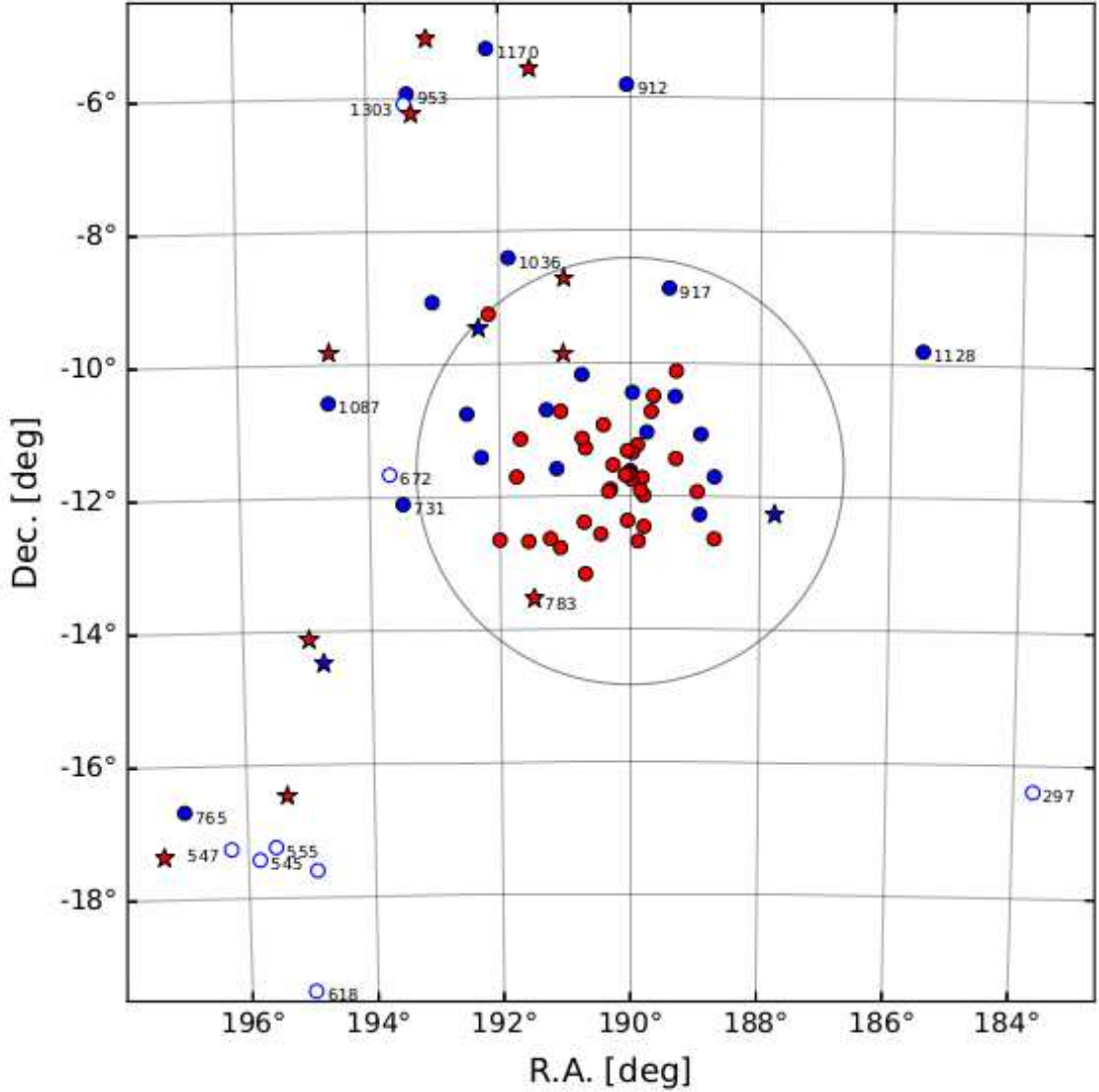

**Figure 5**: Distribution of the LV dwarfs around the Sombrero galaxy. The galaxies of early types (Sph, dE) and late types (Irr, Im, Tr, BCD, Sm, Sdm) are shown by red and blue symbols, respectively. Supposed foreground galaxies with $D < 7.3$ Mpc are indicated with open circles. The dwarf galaxies we discovered are shown as blue and red asterisks according to their morphological type.

Carlsten et al. (2022) used images obtained with the 3.6 meter Canadian-French-Hawaii Telescope to discover 11 new satellites of Sombrero in a region with a radius of $0.6°$ or $\sim 100$ kpc. The belonging of these dwarfs to the Sombrero group was confirmed by measuring their distances using the surface brightness fluctuation (sbf) method. The most significant addition to the number of dwarf galaxies in the Sombrero neighborhood was given by the study of Crosby et al. (2024a). Using images obtained with the Subaru Supreme-Cam, they discovered 40 new potential satellites of Sombrero within a circle of $\sim 2°$ radius. Surface photometry was performed for these galaxies and integral properties of the galaxies were determined. Of these 40 dwarfs, 27 were classified as satellites of Sombrero with high confidence.

Looking at a wider region, Karachentsev et al. (2020a), identified 15 supposed satellites of the Sombrero with measured radial velocities. Their average projected separation relative to the Sombrero is 431 kpc, and their radial velocity dispersion is 204 km s$^{-1}$. Applying relations (1) and (3) to this ensemble of satellites yields an estimate of the total mass of the group of $M_T = (15.5 \pm 4.9) \times 10^{12} M_\odot$ and a virial radius of $R_v = 538$ kpc or $3.23°$.

Using the DR10 data of DESI Legacy Imaging Surveys, we undertook searches for new satellites of Sombrero in the region : RA= $[183, 197]°$, Dec= $[-5, -19]°$, and found 13 dwarf galaxies presented in Table 2. We also re-examined images of 27 dwarf galaxies from the Crosby et al. (2024a) list, identified as Sombrero satellites with high confidence. In our estimation, 25 of them look like real members of the Sombrero group. But we considered two galaxies as belonging to the distant background: UGCA 287 due to its estimated distance of $D_{TF} = 20.5$ Mpc (Karachentsev and Nasonova 2013), and the galaxy dw 1241-1008 due to its texture.

It should be noted that the Sombrero group is located near the equator of the Local Supercluster, projecting onto the Virgo Southern Extension filament (Kourkchi and Tully 2017, Tully 1982). Many galaxies in this filament have radial velocities close to that of Sombrero ($V_{\rm LG} = 912$) km s$^{-1}$, but are located at distances of 15–20 Mpc, typical for members of the Virgo cluster. The presence of the rich background makes it difficult to single out Sombrero's satellites.

We selected from the LV database (http://www/lv/lvgdb) the galaxies with distance estimates $D < 12$ Mpc, which are located in the square of $14°$ by $14°$ around Sombrero. This region extends over $2R_v$, encompassing the bulk of the galaxy halo. A list of 68 of these galaxies is presented in Table 4. Its columns contain: (1) — galaxy name as it indicated in UNGC; (2,3) — equatorial coordinates, J2000.0, in degrees; (4) — galaxy distance in Mpc; (5) — method used to determine the distance; (6) — radial velocity in the Local Group rest frame (in km s$^{-1}$); (7) — morphological type; (8) — references to the distance estimates. All radial velocities are taken from the HyperLEDA (Makarov et al. 2014) with the addition of a new radial velocity for UGCA 287 from Crosby et al. (2024b).

The distribution of the LV galaxies in the considered sky region is presented in Figure 5. The galaxies of early types (Sph, dE) and late types (Irr, Im, Tr, BCD, S) are shown by red and blue symbols, respectively. Supposed foreground galaxies with $D < 7.3$ Mpc are indicated with open circles. The dwarf galaxies we discovered are shown as blue and red asterisks according to their morphological type. The large circle of $3.23°$ radius corresponds to $R_v$. The radial velocities of galaxies located outside the dense central zone are indicated by numbers (in km s$^{-1}$).

The distribution of early-type dwarf galaxies shows the expected stronger concentration towards the host galaxy. The distribution of suspected satellites appears somewhat asymmetric. On the right side of this map, outside the virial circle, only one probable satellite of Sombrero has been detected. It is KKSG 27 with $V_{\rm LG} = 1128$ km s$^{-1}$.

In the lower left corner of Fig. 5, a group of foreground galaxies stands out with a typical radial velocity of $V_{\rm LG} \sim 600$ km s$^{-1}$ and a distance of $D \sim 6$ Mpc. This association of dwarf galaxies around DDO 161 likely includes also the two spheroidal dwarfs, Dw 1301-1627 and Dw 1309-1721, we discovered. Three more new Sph dwarfs are located in the upper left part of the figure near DDO 148 ($V_{\rm LG} = 1170$ km s$^{-1}$). It remains unclear whether these galaxies are members of the Sombrero group or belong to outskirts of the Virgo cluster. Obviously, for a better understanding the structure and kinematics of the Sombrero group, it is necessary to measure more distances and radial velocities of galaxies in the considered sky region.

## 5. CONCLUSIONS

We searched for new dwarf companions around 20 southern galaxies of the Local Volume having distances within (4 - 11) Mpc and $\log(L_K/L_\odot)$ luminosities in the range of (8.1–11.3). For half of the host galaxies, no new companions were detected in an area inside of two virial radii. We found eight new companion candidates for the spiral galaxy NGC 6744, which is the most extended galaxy in the LV. The population of NGC 6744 group currently includes 25 candidate members. Based on seven supposed satellites of NGC 6744 with measured radial velocities, the total mass of the group is estimated to be $(1.88 \pm 0.71) \times 10^{12} M_\odot$. The accuracy of determining the total mass of this group may be noticeably increased when measuring radial velocities of its other members.

Our search for dwarf galaxies in a 14° by 14° square centered on M 104 (Sombrero) has resulted in 13 new dwarfs. Most of these are likely associated with the Sombrero itself and with a foreground diffuse group of dwarf galaxies around DDO 161. We also rediscovered 25 of the 27 objects recently found by Crosby et al. (2024b) as members of the Sombrero group with high confidence. A deep 21-cm survey of this region with sufficiently high angular resolution, likes the Wallaby survey (Koribalski et al. 2020), could elucidate the structure and kinematics of the galaxy ensemble in this intricate sky region adjacent to the Virgo Cluster.

## ACKNOWLEDGEMENTS

We are grateful to an anonymous referee for constructive comments, in particular for drawing our attention to the peculiarity of the configuration of satellites around the NGC 6744. This work has made use of DESI Legacy Imaging Surveys data, the HyperLEDA, and the revised version of the Local Volume galaxies database. This work was supported by the Russian Science Foundation grant 24–12–00277.

Table 1: Southern host galaxies in the Local Volume.

| Name | RA, Dec | $\log(L_K/L_\odot)$ | D | meth | N | deg |
|---|---|---|---|---|---|---|
| (1) | (2) | (3) | (4) | (5) | (6) | (7) |
| NGC 0024 | 02.48−24.96 | 9.48 | 7.31 | TRGB | 1 | 3.1 |
| NGC 0045 | 03.52−23.18 | 9.33 | 6.64 | TRGB | 2 | 3.0 |
| NGC 0625 | 23.77−41.44 | 8.96 | 4.02 | TRGB | 1 | 3.7 |
| ESO 300-014 | 47.41−41.03 | 9.30 | 9.80 | TF | 0 | 2.0 |
| NGC 1313 | 49.56−66.50 | 9.57 | 4.31 | TRGB | 1 | 5.6 |
| NGC 1311 | 50.03−52.18 | 8.43 | 5.55 | TRGB | 0 | 1.8 |
| NGC 1592 | 67.43−27.41 | 8.17 | 9.10 | TF | 0 | 0.9 |
| NGC 1637 | 70.37−02.86 | 10.07 | 9.29 | SN | 0 | 3.8 |
| NGC 1800 | 76.61−31.95 | 9.04 | 8.00 | TF | 0 | 2.0 |
| NGC 2188 | 92.54−34.11 | 9.37 | 8.39 | TRGB | 0 | 2.5 |
| NGC 3037 | 147.85−27.01 | 8.90 | 10.28 | NAM | 0 | 1.4 |
| NGC 3621 | 169.57−32.81 | 10.34 | 6.64 | TRGB | 3 | 6.5 |
| NGC 4594 | 190.00−11.62 | 11.32 | 9.55 | TRGB | 13 | 7.0 |
| NGC 5068 | 199.73−21.04 | 9.73 | 5.15 | TRGB | 5 | 5.3 |
| NGC 6744 | 287.44−63.86 | 10.91 | 9.51 | TRGB | 8 | 7.1 |
| IC 5052 | 313.03−69.20 | 9.27 | 5.50 | TRGB | 0 | 3.5 |
| NGC 7090 | 324.12−54.56 | 10.03 | 9.51 | TRGB | 1 | 3.6 |
| NGC 7424 | 344.33−41.07 | 10.08 | 9.86 | TF | 5 | 3.6 |
| IC 5332 | 353.62−36.10 | 9.74 | 9.01 | TRGB | 0 | 3.1 |
| NGC 7713 | 354.06−37.94 | 9.46 | 8.05 | TRGB | 0 | 2.8 |

Table 2: New candidate dwarf galaxies in the Local Volume .

| Name | RA (2000.0) Dec | $a'$ | $b/a$ | $B$ | Type |
|---|---|---|---|---|---|
| (1) | (2) | (3) | (4) | (5) | (6) |
| ESO 472-015 | 00 06 48.4 −24 56 41 | 1.53 | 0.56 | 16.24 | Im |
| PGC 73354 | 00 17 01.4 −23 42 30 | 0.77 | 0.57 | 16.89 | BCD |
| PGC 01242 | 00 19 11.5 −22 40 06 | 3.06 | 0.51 | 15.38 | BCD |
| Dw 0128-4116 | 01 28 43.0 −41 16 34 | 0.68 | 0.52 | 17.68 | Irr |
| Dw 0312-6643 | 03 12 21.1 −66 43 16 | 0.64 | 0.63 | 21.0 | Sph |
| Dw 1105-3041 | 11 05 47.5 −30 41 19 | 0.77 | 0.84 | 20.0 | Sph |
| KK 101 | 11 16 44.6 −32 39 04 | 1.06 | 0.72 | 17.56 | Sph |
| Dw 1128-3225 | 11 28 58.1 −32 25 23 | 0.47 | 0.55 | 20.5 | Irr |
| Dw 1231-1216 | 12 31 06.0 −12 16 12 | 0.72 | 0.73 | 18.7 | Tr |
| Dw 1244-0842 | 12 44 01.9 −08 42 47 | 0.56 | 0.58 | 19.6 | Sph |
| Dw 1244-0951 | 12 44 05.0 −09 51 00 | 0.24 | 0.75 | 21.8 | Sph |
| Dw 1245-1332 | 12 45 54.5 −13 32 31 | 0.87 | 0.70 | 18.8 | Sph |
| Dw 1246-0533 | 12 46 07.2 −05 33 25 | 0.75 | 0.89 | 18.8 | Sph |
| Dw 1249-0928 | 12 49 15.1 −09 28 08 | 0.51 | 0.60 | 21.5 | Tr |
| Dw 1252-0506 | 12 52 20.9 −05 06 18 | 0.80 | 0.79 | 19.0 | Sph |
| Dw 1253-0613 | 12 53 16.1 −06 13 55 | 0.91 | 0.95 | 18.6 | Sph |
| Dw 1258-0948 | 12 58 24.0 −09 48 54 | 0.53 | 0.68 | 21.4 | Sph |
| Dw 1259-1429 | 12 59 01.4 −14 29 02 | 0.66 | 0.68 | 19.1 | Sph |
| Dw 1259-1407 | 12 59 54.2 −14 07 23 | 0.61 | 0.90 | 20.7 | Irr |
| Dw 1301-1627 | 13 01 28.1 −16 27 40 | 0.76 | 0.61 | 18.9 | Sph |
| Dw 1309-1721 | 13 09 18.0 −17 21 54 | 0.52 | 0.78 | 19.9 | Sph |
| KK 186 | 13 09 45.6 −23 32 35 | 1.25 | 0.65 | 17.40 | Tr |
| Dw 1314-1817 | 13 14 35.0 −18 17 20 | 0.83 | 0.81 | 18.00 | Irr |
| Dw 1321-2202 | 13 21 04.1 −22 01 59 | 0.84 | 0.74 | 18.06 | Sph |
| Dw 1321-1935 | 13 21 59.0 −19 35 31 | 0.89 | 0.94 | 18.62 | Sph |
| Dw 1324-2017 | 13 24 54.2 −20 17 53 | 0.41 | 0.70 | 20.5 | Irr |
| Dw 1844-6232 | 18 44 54.5 −62 32 49 | 0.46 | 0.86 | 20.04 | Sph |
| Dw 1851-6400 | 18 51 10.3 −64 00 29 | 0.53 | 0.90 | 18.28 | Tr |
| Dw 1851-6511 | 18 51 39.1 −65 11 38 | 0.56 | 0.93 | 21.5 | Sph |
| Dw 1851-6356 | 18 51 42.2 −63 56 06 | 0.28 | 0.85 | 21.8 | Sph |
| Dw 1851-6355 | 18 51 43.0 −63 55 12 | 0.38 | 0.80 | 21.0 | Sph |
| Dw 1903-6405 | 19 03 39.4 −64 05 56 | 0.32 | 0.81 | 21.7 | Sph |
| KKs 73 | 19 21 46.2 −60 41 00 | 1.24 | 0.64 | 16.90 | Irr |
| KKs 74 | 19 27 43.2 −61 00 38 | 1.11 | 0.41 | 17.60 | Tr |
| Dw 2140-5544 | 21 40 59.3 −55 44 49 | 0.88 | 0.61 | 18.90 | Sph |
| ESO 346-007 | 22 53 23.7 −38 47 59 | 1.60 | 0.68 | 15.06 | BCD |
| Dw 2256-4107 | 22 56 13.9 −41 07 43 | 3.00 | 0.38 | 17.0 | Tr |
| Dw 2256-4250 | 22 56 38.2 −42 50 38 | 1.42 | 0.68 | 17.2 | Irr |
| ESO 290-028 | 22 57 09.2 −42 48 18 | 5.63 | 0.10 | 14.52 | Sdm |
| PGC 580285 | 23 10 38.1 −40 59 11 | 0.83 | 0.76 | 17.51 | Irr |

Table 3: Suggested members of the NGC 6744 galaxy group.

| Name | SGL | SGB | Type | $D$ | meth | $R_p$ | $V_{LG}$ | $M_T$ |
|---|---|---|---|---|---|---|---|---|
| | deg | deg | | Mpc | | kpc | km s$^{-1}$ | $10^{12} M_\odot$ |
| (1) | (2) | (3) | (4) | (5) | (6) | (7) | (8) | (9) |
| Dw 1844-6232 | 205.261 | 11.719 | Sph | 9.51 | mem | 509 | - | - |
| NGC 6684 | 205.807 | 9.110 | Sa | 10.23 | TRGB | 435 | 720 | 0.10 |
| Dw 1851-6400 | 206.022 | 10.280 | Tr | 9.51 | mem | 345 | - | - |
| Dw 1851-6356 | 206.080 | 10.353 | Sph | 9.51 | mem | 335 | - | - |
| Dw 1851-6355 | 206.082 | 10.369 | Sph | 9.51 | mem | 335 | - | - |
| Dw 1851-6511 | 206.093 | 9.095 | Sph | 9.51 | mem | 396 | - | - |
| ESO 104-022 | 206.523 | 9.480 | Irr | 8.95 | TRGB | 302 | 654 | 0.96 |
| dw 1859-64 | 206.943 | 10.246 | Irr | 9.51 | mem | 193 | - | - |
| dw 1901-63 | 207.192 | 11.021 | Sph | 9.51 | mem | 184 | - | - |
| Dw 1903-6405 | 207.409 | 10.175 | Sph | 9.51 | mem | 120 | - | - |
| N 6744dwTBGa | 207.700 | 10.990 | Sph | 7.61 | sbf | 121 | - | - |
| KKs 70 | 207.784 | 10.292 | Irr | 7.31 | sbf | 55 | - | - |
| dw 1907-63 | 207.842 | 10.549 | Sph | 7.99 | sbf | 51 | - | - |
| KKs 71 | 207.993 | 10.515 | Irr | 9.51 | mem | 28 | - | - |
| EDwC1 | 208.042 | 10.558 | Sph | 9.51 | mem | 31 | - | - |
| NGC 6744 | 208.101 | 10.382 | Sc | 9.51 | TRGB | 0 | 706 | 0 |
| ESO 104-044 | 208.254 | 10.010 | Irr | 9.73 | TRGB | 66 | 614 | 0.65 |
| KKs 72 | 208.394 | 10.364 | Irr | 9.51 | mem | 63 | - | - |
| N 6744dwTBGb | 208.453 | 10.553 | Tr | 9.51 | mem | 65 | - | - |
| IC 4824 | 208.630 | 12.119 | Im | 10.30 | NAM | 301 | 811 | 3.91 |
| AM1909-613 | 208.722 | 12.287 | Sm | 9.51 | mem | 333 | 821 | 5.20 |
| ESO 141-042 | 208.955 | 11.819 | Sm | 9.70 | NAM | 332 | 769 | 1.55 |
| KKs 73 | 209.811 | 13.419 | Irr | 9.51 | mem | 578 | - | - |
| KKs 74 | 210.507 | 13.002 | Tr | 9.51 | mem | 590 | - | - |
| IC 4870 | 210.838 | 8.086 | BCD | 8.51 | TRGB | 592 | 740 | 0.81 |

Table 4: LV galaxies within RA=[183.0 −197.0], Dec=[−5.0, −19.0]

| Name | RA, ° | Dec, ° | D, Mpc | meth | $V_{LG}$ km s$^{-1}$ | Type | Ref. |
|---|---|---|---|---|---|---|---|
| (1) | (2) | (3) | (4) | (5) | (6) | (7) | (8) |
| Corvus A | 183.690 | −16.397 | 3.48 | TRGB | 297 | Tr | [1] |
| KKSG 27 | 185.524 | −9.800 | 9.08 | bTF | 1128 | Im | [2] |
| dw 1234-1238 | 188.704 | −12.640 | 9.55 | mem | – | Sph | – |
| dw 1234-1142 | 188.704 | −11.707 | 9.55 | mem | – | Tr | – |
| LV J1235-1104 | 188.914 | −11.067 | 10.00 | TF | 1003 | BCD | [3] |
| dw 1235-1216 | 188.929 | −12.273 | 9.55 | mem | – | Irr | – |
| dw 1235-1155 | 188.967 | −11.931 | 9.55 | mem | – | Sph | – |
| dw 1237-1125 | 189.298 | −11.433 | 7.51 | sbf | – | dE | [4] |
| dw 1237-1006 | 189.292 | −10.116 | 9.55 | mem | – | Sph | – |
| KKSG 29 | 189.309 | −10.498 | 9.82 | TRGB | 562 | Irr | [5] |
| KKSG 30 | 189.400 | −8.867 | 9.73 | TRGB | 917 | Irr | [6] |
| KKSG 31 | 189.640 | −10.490 | 9.55 | mem | – | Sph | – |
| dw 1238-1043 | 189.675 | −10.725 | 9.55 | mem | – | Sph | – |
| dw 1238-1102 | 189.743 | −11.04 | 9.55 | mem | – | Irr | – |
| dw 1239-1227 | 189.788 | −12.454 | 9.55 | mem | – | Sph | – |
| dw 1239-1159 | 189.788 | −11.987 | 11.33 | sbf | – | Sph | [4] |
| NGC 4594DW1 | 189.814 | −11.719 | 9.42 | sbf | 1171 | dE | [7] |
| dw 1239-1154 | 189.842 | −11.907 | 9.55 | mem | – | Sph | – |
| dw 1239-1240 | 189.875 | −12.675 | 9.55 | mem | – | Sph | – |
| NGC 4594-DGSAT-3 | 189.887 | −11.227 | 7.87 | sbf | – | Sph | [4] |
| dw 1239-1026 | 189.962 | −10.438 | 9.55 | mem | – | Irr | – |
| Sombrero DwA | 189.965 | −11.341 | 9.71 | sbf | – | Sph | [4] |
| KKSG 32 | 189.979 | −11.747 | 9.00 | sbf | – | Sph | [4] |
| NGC 4594 | 189.996 | −11.623 | 9.55 | TRGB | 894 | S0a | [8] |
| SUCD 1 | 190.013 | −11.668 | 9.55 | mem | 1109 | dE | – |
| KKSG 33 | 190.037 | −12.365 | 9.55 | mem | – | Sph | – |
| dw 1240-1118 | 190.039 | −11.314 | 8,79 | sbf | – | dE | [4] |
| NGC 4597 | 190.054 | −5.799 | 10.10 | TF | 912 | Sm | [9] |
| dw 1240-1140 | 190.073 | −11.679 | 9.55 | mem | 1097 | dE | – |
| dw 1241-1131 | 190.262 | −11.529 | 9.55 | mem | – | Sph | – |
| Sombrero DwB | 190.300 | −11.892 | 11.19 | sbf | – | Sph | [4] |
| KKSG 34 | 190.329 | −11.928 | 9.02 | sbf | – | Sph | [4] |
| dw 1241-1055 | 190.408 | −10.926 | 9.55 | mem | – | Sph | – |
| dw 1241-1234 | 190.450 | −12.570 | 9.55 | mem | – | Sph | – |
| dw 1242-1116 | 190.683 | −11.274 | 9.55 | mem | – | Sph | – |
| dw 1242-1309 | 190.688 | −13.166 | 9.55 | mem | – | Sph | – |
| PGC 042730 | 190.704 | −12.391 | 9.55 | mem | 828 | dE | – |
| dw 1242-1107 | 190.733 | −11.128 | 9.55 | mem | – | Sph | – |
| dw 1242-1010 | 190.738 | −10.169 | 9.55 | mem | – | Irr | – |

**Table 4**: LV galaxies within RA=[183.0 −197.0], Dec=[−5.0, −19.0]

| Name | RA, ° | Dec, ° | $D$, Mpc | meth | $V_{\rm LG}$ km s$^{-1}$ | Type | Ref. |
|---|---|---|---|---|---|---|---|
| (1) | (2) | (3) | (4) | (5) | (6) | (7) | (8) |
| Dw 1244-0842 | 191.008 | −8.713 | 9.55 | mem | – | Sph | – |
| dw 1244-1043 | 191.062 | −10.722 | 9.55 | mem | – | Sph | – |
| dw 1244-1246 | 191.071 | −12.773 | 9.55 | mem | – | Sph | – |
| dw 1244-1135 | 191.125 | −11.585 | 9.55 | mem | – | Irr | – |
| dw 1244-1238 | 191.225 | −12.636 | 9.55 | mem | – | Sph | – |
| dw 1245-1041 | 191.279 | −10.699 | 9.55 | mem | – | Tr | – |
| Dw 1245-1332 | 191.477 | −13.542 | 9.55 | mem | 783 | Sph | – |
| dw 1246-1240 | 191.558 | −12.679 | 9.55 | mem | – | Sph | – |
| dw 1246-1108 | 191.679 | −11.138 | 9.55 | mem | – | Sph | – |
| dw 1246-1142 | 191.742 | −11.707 | 9.55 | mem | - | Sph | – |
| Dw 1247-0824 | 191.854 | −8.408 | 9.55 | mem | 1036 | BCD | – |
| KKSG 37 | 192.004 | −12.655 | 9.55 | mem | – | Sph | – |
| Dw 1248-0915 | 192.160 | −9.256 | 9.55 | mem | – | Sph | – |
| DDO 148 | 192.180 | −5.254 | 9.00 | TF | 1170 | Sm | [9] |
| NGC 4700 | 192.282 | −11.411 | 7.30 | TF | 1222 | Sdm | [9] |
| PGC 974136 | 192.498 | −10.757 | 9.55 | mem | – | Im | – |
| Dw 1252-0904 | 193.013 | −9.075 | 9.55 | mem | – | Irr | – |
| KKSG 38 | 193.382 | −5.928 | 10.40 | TF | 953 | Irr | [9] |
| KKSG 39 | 193.425 | −6.084 | 6.80 | bTF | 1303 | Irr | [2] |
| DDO 153 | 193.490 | −12.1086 | 9.84 | TRGB | 731 | Im | [6] |
| PGC 963198 | 193.691 | −11.661 | 6.15 | TF | 672 | Im | [2] |
| Dw 1258-0948 | 194.600 | −9.815 | 9.55 | mem | – | Sph | – |
| MCG-02-33-75 | 194.618 | −10.577 | 8.70 | TF | 1087 | Sm | [9] |
| Dw 1259-1735 | 194.918 | −17.596 | 6.03 | mem | – | Irr | – |
| KK 176 | 194.985 | −19.413 | 7.28 | TRGB | 618 | Irr | [5] |
| UGCA 319 | 195.560 | −17.238 | 5.75 | TRGB | 555 | Irr | [5] |
| DDO 161 | 195.820 | −17.423 | 6.03 | TRGB | 545 | Sm | [5] |
| PGC 886203 | 196.266 | −17.258 | 5.61 | NAM | 547 | Irr | [2] |
| MCG-03-34-2 | 196.986 | −16.689 | 7.90 | TF | 765 | BCD | [9] |

**Notes**: [1] — (Jones et al. 2024b); [2] — present paper;
[3] — (Kashibadze et al. 2018); [4] — (Carlsten et al. 2022);
[5] — (Karachentsev et al. 2018); [6] — (Karachentsev et al. 2020a);
[7] — (Carlsten et al. 2020); [8] — (McQuinn et al. 2016);
[9] — (Karachentsev and Nasonova 2013);